\documentclass[12pt]{iopart}
\usepackage{epsfig}

\begin{document}

\title[Jet Propagation and Mach Cones in (3+1)d Ideal Hydrodynamics]{Jet Propagation and Mach Cones in (3+1)d Ideal Hydrodynamics}

\author{Barbara Betz$^{1,2}$, Miklos Gyulassy$^3$, Dirk H. Rischke$^{1,4}$, Horst St\"ocker$^{4,5}$, Giorgio Torrieri$^1$}
\address{$^1$Institut f\"ur Theoretische Physik, J.W. Goethe-Universit\"at, Frankfurt, Germany}
\address{$^2$Helmholtz Research School, Universit\"at Frankfurt, GSI and FIAS, Germany}
\address{$^3$Department of Physics, Columbia University, New York, USA}
\address{$^4$Frankfurt Institute for Advanced Studies (FIAS), Frankfurt, Germany}
\address{$^5$Gesellschaft f\"ur Schwerionenforschung, GSI, Darmstadt, Germany}
\ead{betz@th.physik.uni-frankfurt.de}

\vspace*{-0.1cm}
\begin{abstract}
The observation of jet quenching and associated away--side Mach cone--like correlations 
at RHIC provide powerful ``external'' probes of the sQGP produced in A+A reactions 
\cite{MachCone1}, but it simultaneously raises the question where the jet energy was deposited. 
The nearly perfect bulk fluidity observed via elliptic flow suggests that Mach cone--like 
correlations may also be due to rapid local equilibration in the wake of penetrating jets. 
Multi-particle correlations lend further support to this possibility 
\cite{MachCone2}. However, a combined study of energy deposition and fluid response is needed. 
We solve numerically 3--dimensional ideal
hydrodynamical equations to compute the flow correlation patterns resulting 
from a variety of possible energy-momentum deposition models. 
Mach--cone correlations are shown to depend critically on the energy and momentum deposition mechanisms. They 
only survive for a special limited class of energy--momentum loss models,
which assume significantly less longitudinal momentum loss than energy loss per unit length. 
We conclude that the correct interpretation of away--side jet correlations will require 
improved understanding and independent experimental constraints on the jet energy--momentum 
loss to fluid couplings.
\end{abstract}

\vspace*{-0.4cm}
\section{Introduction}

One of the major findings at the Relativistic Heavy Ion Collider (RHIC) is the suppression of highly energetic particles
\cite{MachCone1}. Two-- and three--particle correlations of jet--associated intermediate--$p_\bot$ particles provide an important
test of the response of the medium to the details of the jet--quenching dynamics \cite{MachCone2}.

The observation of strong flow \cite{Kolb:2003dz} suggests the possibility that the energy lost is quickly thermalized and incorporated in the 
local hydrodynamical flow. For a quantitative comparison to data a detailed model of both energy and momentum deposition 
coupled with a relativistic fluid model is needed \cite{Casalderrey,EnergyLoss,MediumResponse}.

We solve numerically $(3+1)$d ideal hydrodynamics \cite{Rischke:1995pe}, including a Bag Model Equation of State (EoS) 
with a critical temperature of $169$~MeV to study the interactions
of the jet with a medium for different energy-- and momentum--deposition scenarios and to compute the flow--correlation 
patterns for the different energy--momentum--deposition models. Our focus is to study how 
hydrodynamical flow profiles (such as Mach cones), defined in configuration space, translate into momentum--space
correlation functions via freeze--out. For this, we investigate the simplest situation of a uniform
medium.

\section{Jets in Ideal Hydrodynamics}

\begin{figure*}[t]
\begin{minipage}[b]{5cm}
\hspace*{1.5cm}
\includegraphics[scale=0.5]{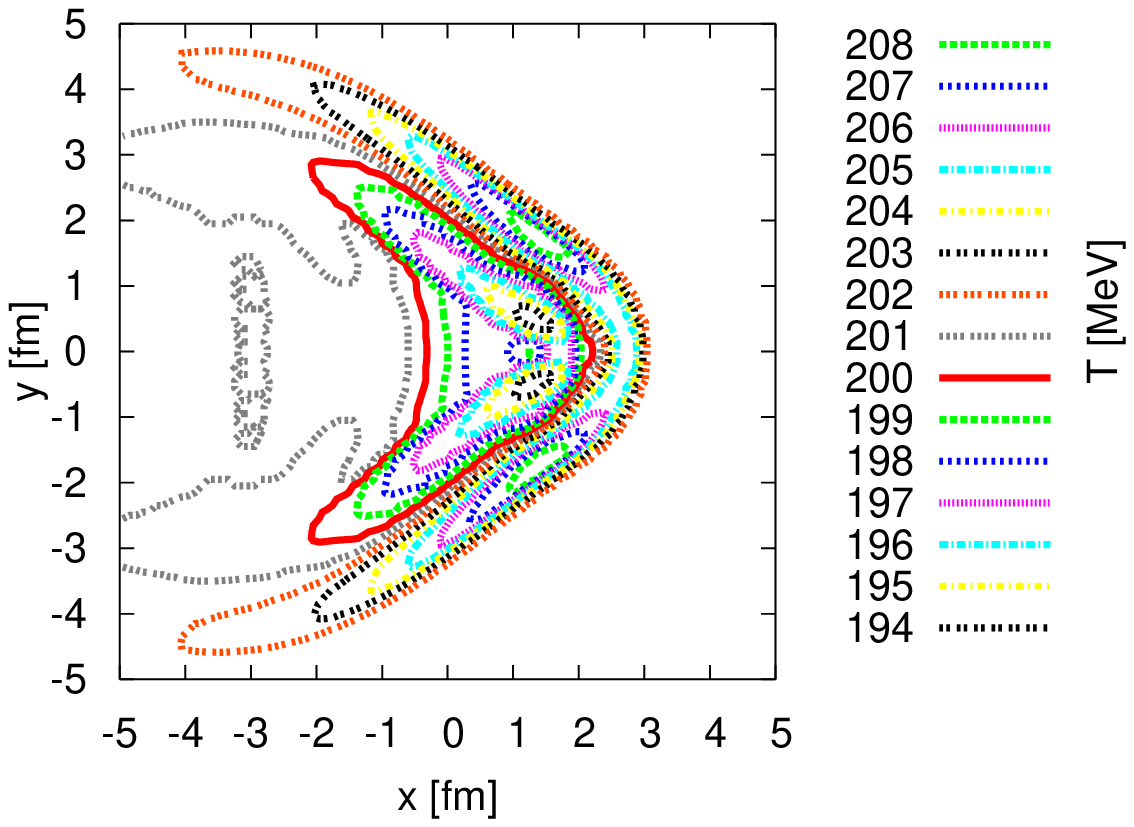}
\end{minipage} 
\hspace*{2.5cm}
\begin{minipage}[b]{5cm}
\includegraphics[scale=0.5]{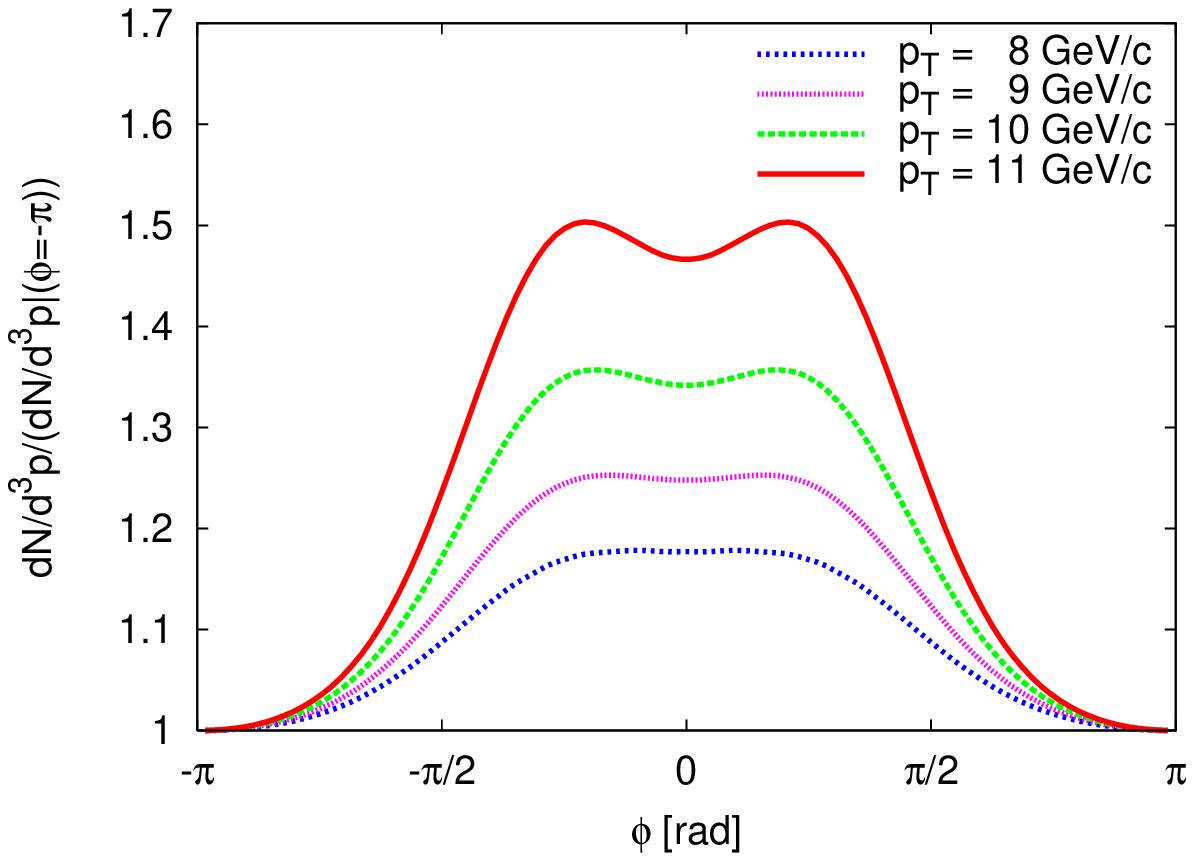}
\end{minipage} 
\vspace*{-0.2cm}
\hspace*{-0.5cm}\caption{(Left panel) Temperature pattern after a hydrodynamical evolution of $t=7.2$~fm/c, assuming 
a pure energy loss of $dE/dx = 1.4$~GeV/fm of a jet moving with $v_{\rm{jet}}=0.99{\rm c}$ along
the $x$--axis. (Right panel) Jet--signal strength determined after isochronous
freeze--out for different $p_\bot$ values. }
\vspace*{-0.2cm}
\label{fig1}
\end{figure*}

In ideal hydrodynamics, the energy--momentum tensor is locally conserved.
Adding a jet to the system, an extended set of equations including a source term $S^\nu$ has to be solved numerically,
\begin{eqnarray}
\partial_\mu T^{\mu\nu} =S^\nu\;.
\end{eqnarray} 
In this work, we will apply a source term
\begin{eqnarray}
\label{source}
S^\nu = \int\limits_{\tau_i}^{\tau_f}d\tau \frac{dP^\nu}{d\tau}\delta^{(4)}\left( x^\mu - x^\mu(\tau) \right),
\end{eqnarray}
and assume a constant energy and momentum loss rate $dP^\nu/d\tau = (dE/d\tau,d\vec{M}/d\tau)$ 
along the trajectory of a jet $x^\mu (\tau) = x_0^\mu + u^\mu_{\rm jet}\tau$,
which moves with nearly the speed of light $(v_{\rm jet}=0.99{\rm c})$ through a homogeneous, non--expanding background.
We terminate the energy--momentum deposition of the jet after $5.6$~fm/c of the hydrodynamical evolution.
Additionally, we assume that the near--side jet contribution
to the correlation function is not affected by the medium, although the 
observation of the ridge makes this assumption far from guaranteed.

Using the Cooper--Frye formula at midrapidity for pions,
we perform an isochronous freeze--out after an evolution of $t=7.2$~fm/c and determine the jet--signal strength
by calculating the triple differential momentum distribution in jet direction (which is the direction of the away--side parton 
propagation) and normalizing to the distribution in direction opposite to the
jet. Additionally, we determine the azimuthal two--particle correlations $1/N dN/d\phi dy\vert (y=0)$ for the different deposition mechanisms.

\section{Jet--Deposition Mechanisms}

\begin{figure*}[t]
\begin{minipage}[b]{4cm}
\hspace*{1.5cm}
\includegraphics[scale=0.5]{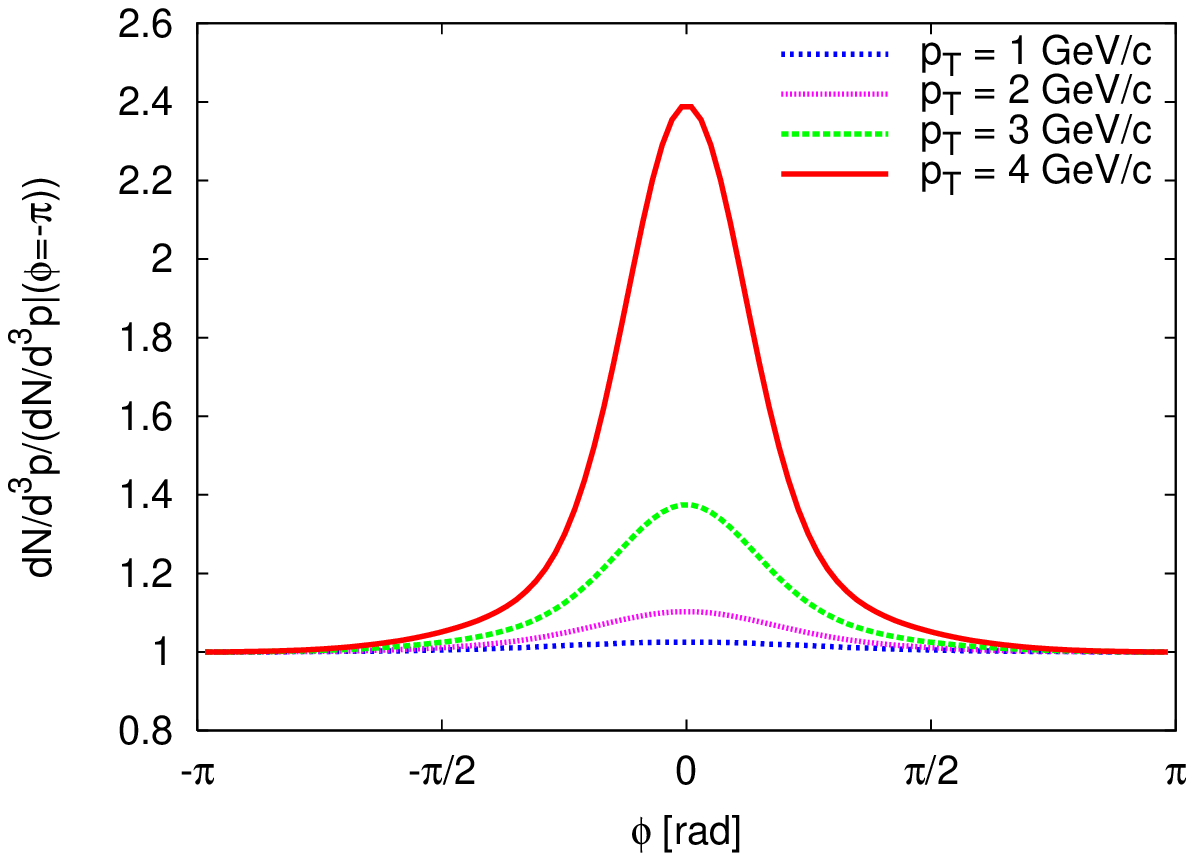}
\end{minipage} 
\begin{minipage}[b]{4cm}
\hspace*{5.3cm}
\includegraphics[scale=0.3]{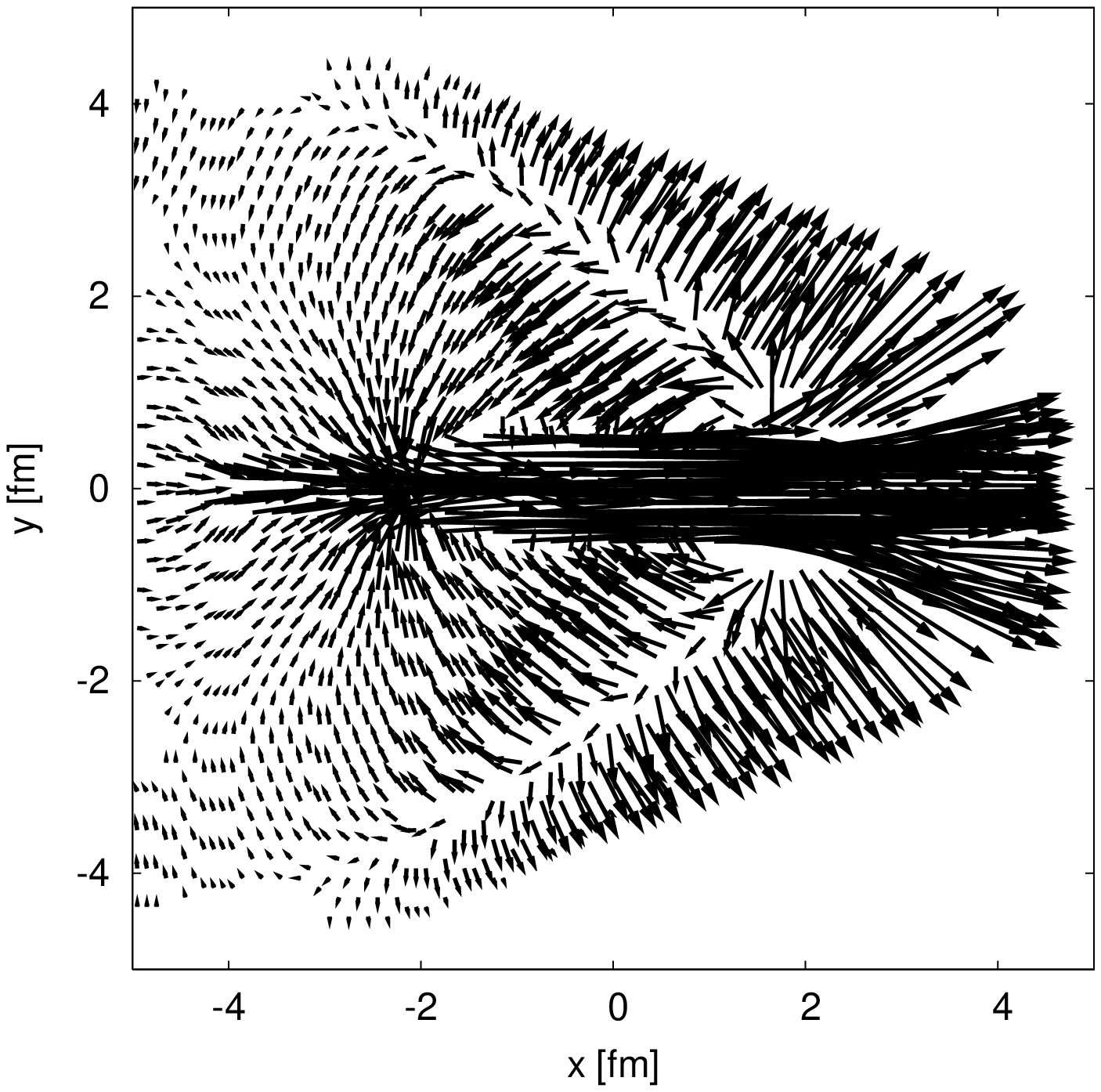}
\end{minipage}
\caption{Jet--signal strength for different $p_\bot$ values (left panel) and momentum distribution (right panel) after a 
hydrodynamical evolution of $t=7.2$~fm/c, assuming a pure momentum deposition of $dM/dx = 1.4$~GeV/fm of a jet moving with $v_{\rm{jet}}=0.99{\rm c}$ along
the $x$--axis.}
\label{fig3} 
\vspace*{-0.2cm}
\end{figure*}


In a first scenario, we study a source term which describes pure energy deposition, i.e.,\ $dP^\nu/d\tau=(dE/d\tau,\vec{0})$ in Eq.\
(\ref{source}), with $dE/dx = 1.4$~GeV/fm. 
The temperature pattern after $t=7.2$~fm/c (see left panel of 
Fig.\ \ref{fig1}) reveals the formation of a cone--like structure. 
In the jet signal strength (see right panel of Fig.\ \ref{fig1}) 
this cone indeed appears in form of a double-peaked structure, but only if high $p_\bot$ values are selected. 
This is due to the fact that thermal smearing washes out the signal for a high background temperature.
Therefore, using $p_\bot$ cuts similar to the experiment ($3 \leq p_\bot \leq 5$~GeV/fm), no cone--like 
structure emerges in the azimuthal two--particle correlations, but a broad away--side peak, if 
$dE/dx < 9$~GeV/fm. Of course, such away-side peaks will be produced in {\em any} jet-quenching mechanisms
consistent with energy--momentum conservation, and hence its experimental observation is not
enough to show that the jet energy has been locally thermalized. To do this, one might
check if the height of the peak rises exponentially with the associated $p_T$.



As a second scenario, we investigate a source term with pure momentum deposition, i.e.,\ $dP^\nu/d\tau=(0,d\vec{M}/d\tau)$ in Eq.\
(\ref{source}), with $dM/dx = 1.4$~GeV/fm. 
This is justified since partons can be virtual.
In this case, one peak occurs in jet direction (as can be seen from the left panel of Fig.\ \ref{fig3}), 
which is already visible for lower $p_\bot$ values as compared to the first deposition scenario (cf.\ right panel of Fig.\ \ref{fig1}). 
The reason is that a diffusion wake is excited, which is indicated by the strong flow in jet direction (see right panel of Fig.\ 
\ref{fig3}).

\begin{figure*}[t]
\hspace*{5cm}
\includegraphics[scale=0.5]{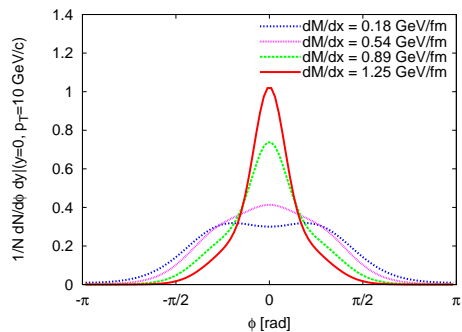} 
\vspace*{-0.4cm}
\caption{Azimuthal two--particle correlation, assuming a jet depositing both energy and momentum, for a jet--energy 
loss of $dE/dx = 1.4$~GeV/fm and different ratios of the totally distributed momentum $M_{\rm jet}$ to the totally distributed energy 
$E_{\rm jet}$ for a value of $p_\bot = 10$~GeV/c.}
\vspace*{-0.6cm}
\label{fig4}
\end{figure*}

The third scenario which we consider is characterized by a source term that describes a combined deposition of energy and momentum
for a jet--energy loss of $dE/dx = 1.4$~GeV/fm and different ratios of the totally distributed momentum $M_{\rm jet}$ to the totally 
distributed energy $E_{\rm jet}$. 
The cone--like shape only emerges for a small jet--momentum loss $dM/dx$ and -- due to the small value 
of the jet--energy loss $dE/dx$ -- for a high $p_\bot$ value (see Fig.\ \ref{fig4}).
For a larger jet--momentum loss, this structure is dissolved (caused by the creation of a diffusion wake) and a peak occurs in 
jet direction.

\section{Summary}
We found that the fluid response to a jet critically depends on the energy--momentum--deposition mechanism. A Mach cone--like pattern
occurs in the azimuthal two--particle correlation only if the longitudinal jet--momentum loss is significantly less
than the jet--energy loss ($dM/dx \ll dE/dx$), since otherwise the diffusion wake kills the Mach cone--like signal.
This result is consistent with Refs.\ \cite{Casalderrey}.
Moreover, applying $p_\bot$ cuts similar to the experimentally used values ($3 \leq p_\bot \leq 5$~GeV/fm), a 
double--peaked conical signal does not emerge in the azimuthal two--particle correlation if $dE/dx < 9$~GeV/fm.

For a correct interpretation of the away--side jet correlations it is necessary to determine a realistic energy--momentum--deposition 
scenario in an expanding medium.

\end{document}